\documentclass{article}
\usepackage{ismir,amsmath,cite,url}
\usepackage{graphicx}
\usepackage{color}
\usepackage{amssymb}
\usepackage{siunitx}
\usepackage{multirow}
\usepackage{booktabs}

\makeatletter
\def\inmod#1{\allowbreak\mkern5mu{\operator@font mod}\,\,#1}
\makeatother

\title{Audio-to-Score Alignment\\using Transposition-invariant Features}

\multauthor
{Andreas Arzt$^1$ \hspace{1cm} Stefan Lattner$^{1,2}$} { \\
$^1$ Institute of Computational Perception, Johannes Kepler University, Linz, Austria\\
$^2$ Sony Computer Science Laboratories (CSL), Paris, France\\
{\tt\small andreas.arzt@jku.at}
}
\def\authorname{Andreas Arzt, Stefan Lattner}

\begin{document}

\maketitle
\begin{abstract}
	Audio-to-score alignment is an important pre-processing step for in-depth analysis of classical music.
	In this paper, we apply novel transposition-invariant audio features to this task.
	These low-dimensional features represent local pitch intervals and are learned in an unsupervised fashion by a gated autoencoder.
	Our results show that the proposed features are indeed fully transposition-invariant and enable accurate alignments between transposed scores and performances.
	Furthermore, they can even outperform widely used features for audio-to-score alignment on `untransposed data', and thus are a viable and more flexible alternative to well-established features for music alignment and matching.

\end{abstract}
\section{Introduction}\label{sec:introduction}

The task of synchronising an audio recording of a music performance and its score has already been studied extensively in the area of intelligent music processing. It forms the basis for multi-modal inter- and intra-document navigation applications \cite{dixon:ismir:2005,FremereyKMC07_DemonstrationSyncplayerSystem_ISMIR,Thomas13_SheetMusicAudio_PhD} as well as for the analysis of music performances, where e.g. aligned pairs of scores and performances are used to extract tempo curves or learn predictive performance models \cite{WidmerDGPT03_Horowitz_AI, grachten:ieeemult:2017}.


Typically, this synchronisation task, known as audio-to-score alignment, is based on a symbolic score representation, e.g. in the form of MIDI or MusicXML. In this paper, we follow the common approach of converting this score representation into a sound file using a software synthesizer. The result is a low-quality rendition of the piece, in which the time of every event is known. Then, for both sequences the same kinds of features are computed, and a sequence alignment algorithm is used to align the audio of the performance to the audio representation of the score, i.e. the problem of audio-to-score alignment is treated as an audio-to-audio alignment task. The output is a mapping, relating all events in the score to time points in the performance audio.  Common features for this task include a variety of chroma-based features\cite{HuDT03_audiomatching_WASPAA,EllisP07_CoverSong_ICASSP,JoderER10_MusicAlignment_ICASSP, Mueller15_FMP_SPRINGER}, features based on the semitone scale\cite{dixon:ismir:2005,arzt:eusipco:2012}, and mel-frequency cepstral coefficients (MFCCs)\cite{grachten:ismir:2013}.

In this paper, we apply novel low-dimensional features to the task of music alignment. The features represent local pitch intervals and are learned in an unsupervised fashion by a gated autoencoder \cite{memisevic2011gradient}. We will demonstrate how these features can be used to synchronise a recording of a performance to a transposed version of its score. Furthermore, as they are only based on a local context, the features can even cope with multiple transpositions within a piece with only minimal additional alignment error, which is not possible at all with common pitch-based feature representations.

The main contributions of this paper are (1) the introduction of novel transposition-invariant features to the task of music synchronisation, (2) an in-depth analysis of their properties in the context of this task, and (3) a direct comparison to chroma features, which are the quasi-standard for this task. A cleaned-up implementation of the code for the gated autoencoder used in this paper is publicly available\footnote{see \url{https://github.com/SonyCSLParis/cgae-invar}}. The paper is structured as follows. In Section \ref{sec:feature_computation}, the features are introduced. Section \ref{sec:alignment_algorithm} briefly describes the alignment algorithm we are using throughout the paper. Then, in Section \ref{sec:experiments_piano} we present detailed experiments on piano music, including a comparison of different feature configurations, results on transposed scores, and a comparison with chroma features. In Section \ref{sec:experiments_orchestra} we discuss the application of the features in the domain of complex orchestral music. Finally, Section \ref{sec:conclusion} gives an outlook on future research directions.

\section{Transposition-invariant Features for Music Synchronisation}\label{sec:feature_computation}

Transposition-invariant methods have been studied extensively in music information retrieval (MIR), for example in the context of music identification \cite{ArztBW12_SymbolicFingerprint_ISMIR}, structure analysis \cite{MuellerC07_Transposition_ISMIR}, and content-based music retrieval \cite{DBLP:conf/cmmr/WaltersRL12,DBLP:journals/tmm/Marolt08,Lemstrm2011TranspositionAT}. However, so far there has been limited success in transposition-invariant audio-to-score alignment. Currently, a typical approach is to first try to identify the transposition, transform the inputs accordingly, and then apply common alignment techniques (see e.g. \cite{sentuerk:etal:jnmr2014}). Another option is to  perform the alignment multiple times, with different transpositions (e.g. the twelve possible transposition options when using chroma-based features) and then select the alignment which produced the least alignment costs (see e.g. \cite{senturk2016compositionIdentification_smc}).

These are cumbersome and error-prone methods. In this paper, we demonstrate how to employ novel transposition-invariant features for the task of score-to-audio alignment, i.e. the features themselves are transposition-invariant. These features have been proposed recently in \cite{lattner2018learning} and their usefulness has been demonstrated for tasks like the detection of repeated (but possibly transposed) motifs, themes and sections in classical music.

The features are learned automatically from audio data in an unsupervised way by a gated autoencoder. The main idea is to try to learn a \emph{relative} representation of the current audio frame, based on a small local context (i.e., n-gram, the previous $n$ frames). During the training process, the gated autoencoder is forced to represent this target frame via its preceding frames in a relative way (i.e. via interval differences between the local context and the target frame).

In the following, we give a more detailed description of how these features are learned. Specifics about the training data we are using in this paper can be found in the respective sections on applying this approach to piano music (Section \ref{sec:experiments_piano}) and orchestral music (Section \ref{sec:experiments_orchestra}).

\subsection{Model}\label{sec:model}
Let $\mathbf{x}_t \in \mathbb{R}^M$ be a vector representing the energy distributed over $M$ frequency bands at time t.
Given a temporal context $\mathbf{x}_{t-n}^{t} = \mathbf{x}_{t-n} \dots \mathbf{x}_{t}$ (i.e. the input) and the next time slice $\mathbf{x}_{t+1}$ (i.e. the target), the goal is to learn a mapping $\mathbf{m}_t$ (i.e. the transposition-invariant feature vector at time t) which does not change when shifting $\mathbf{x}_{t-n}^{t+1}$ up- or downwards in the pitch dimension.

\begin{figure}
\begin{center}
\includegraphics[width=.8\linewidth]{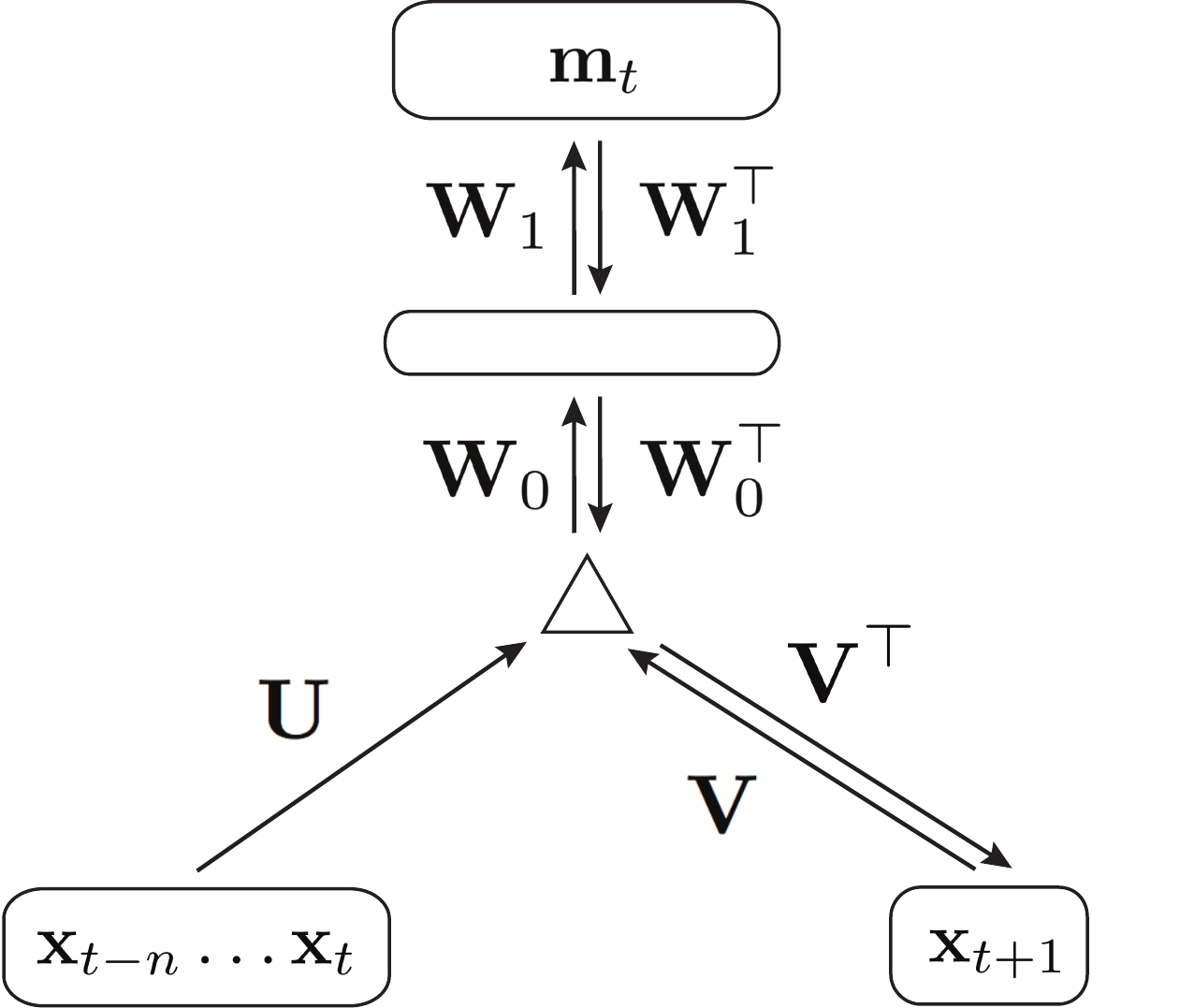}
\caption{Schematic illustration of the gated autoencoder architecture used for feature learning. Double arrows denote weights used for both, inference of the mapping $\mathbf{m}_t$ and the reconstruction of $\mathbf{x}_{t+1}$.}
\label{fig:ga}
\end{center}
\end{figure}

Gated autoencoders (GAEs, see Figure \ref{fig:ga}) are fundamentally different to standard sparse coding models, like denoising autoencoders. GAEs are explicitly designed to learn relations (i.e., covariances) between data pairs by employing an element-wise product in the first layer of the architecture.
In musical sequences, using a GAE for learning relations between pitches in the input and pitches in the target naturally results in representations of musical \emph{intervals}. The intervals are encoded in the latent variables of the GAE as mapping codes $\mathbf{m}_t$ (refer to \cite{lattner2018learning} for more details on interval representations in a GAE).
The goal of the training is to find a mapping for any input/target pair which transforms the input into the given target by applying the represented intervals.
The mapping at time $t$ is calculated as

\begin{equation}\label{eq:gamap}
\mathbf{m}_t = \sigma_h (\mathbf{W}_1 \sigma_h(\mathbf{W}_0(\mathbf{U}\mathbf{x}_{t-n}^{t} \cdot \mathbf{V}\mathbf{x}_{t+1}))),
\end{equation}
where $\mathbf{U}, \mathbf{V}$ and $\mathbf{W}_k$ are weight matrices, and $\sigma_h$ is the hyperbolic tangent non-linearity.
The operator~$\cdot$ (depicted as a triangle in Figure~\ref{fig:ga}) denotes the Hadamard (or element-wise) product of the filter responses $\mathbf{U}\mathbf{x}_{t-n}^{t}$ and $\mathbf{V}\mathbf{x}_{t+1}$, denoted as \emph{factors}.
The target of the GAE can be reconstructed as a function of the input $\mathbf{x}_{t-n}^{t}$ and a mapping $\mathbf{m}_t$:
\begin{equation}\label{recony}
\mathbf{\tilde{x}}_{t+1} = \mathbf{V}^\top (\mathbf{W}_0^\top\mathbf{W}_1^\top \mathbf{m}_t \cdot \mathbf{U}\mathbf{x}_{t-n}^{t}).
\end{equation}

As cost function we use the mean-squared error between the target $\mathbf{x}_{t+1}$ and the target's reconstruction $\mathbf{\tilde{x}}_{t+1}$ as
\begin{equation}\label{eq:cost}
\text{MSE} = \frac{1}{M} \left\| \mathbf{x}_{t+1} - \mathbf{\tilde{x}}_{t+1} \right\|^2.
\end{equation}

\subsection{Training Data Preprocessing}\label{sec:data}

Models are learned directly from audio data, without the need for any annotations. The empirically found preprocessing parameters are as follows. The audio files are re-sampled to 22.05 kHz. We choose a constant-Q transformed spectrogram using a hop size of $448$ ($\sim 20$ms), and Hann windows with different sizes depending on the frequency bin. The range comprises $120$ frequency bins (24 per octave), starting from a minimal frequency of $65.4$ Hz. Each time slice is contrast-normalized to zero mean and unit variance.

\subsection{Training}\label{sec:training}
The model is trained with stochastic gradient descent in order to minimize the cost function (cf. Equation~\ref{eq:cost}) using the training data as described in Section~\ref{sec:data}.
In an altered training procedure introduced below, we randomly transpose the data during training and explicitly aim at transposition invariance of the mapping codes.

\subsubsection{Enforcing Transposition-Invariance}\label{sec:enforc-transp-invar}
As described in Section~\ref{sec:model} the classical GAE training procedure derives a mapping code from an input/target pair, and subsequently penalizes the reconstruction error of the target given the input and the derived mapping code.
Although this procedure naturally tends to lead to similar mapping codes for input target pairs that have the same interval relationships, the training does not explicitly enforce such similarities and consequently the mappings may not be maximally transposition invariant.

Under ideal transposition invariance, by definition the mappings would be identical across different pitch transpositions of an input/target pair. Suppose that a pair $(\mathbf{x}_{t-n}^{t}, \mathbf{x}_{t+1})$ leads to a mapping $\mathbf{m}$ (by Equation~(\ref{eq:gamap})).
Transposition invariance implies that reconstructing a target $\mathbf{x}^{\prime}_{t+1}$ from the pair $({\mathbf{x}^{\prime}}_{t-n}^{t}, \mathbf{m})$ should be as successful as reconstructing $\mathbf{x}_{t+1}$ from the pair $(\mathbf{x}_{t-n}^{t}, \mathbf{m})$ when $({\mathbf{x}^{\prime}}_{t-n}^{t}, \mathbf{x}^{\prime}_{t+1})$ can be obtained from $(\mathbf{x}_{t-n}^{t}, \mathbf{x}_{t+1})$ by a single pitch transposition.

Our altered training procedure explicitly aims to achieve this characteristic of the mapping codes by penalizing the reconstruction error using mappings obtained from transposed input/target pairs.
More formally, we define a transposition function $\textit{shift}(\mathbf{x}, \delta)$, shifting the values (CQT frequency bins) of a vector $\mathbf{x}$ of length $M$ by $\delta$ steps:
\begin{equation}\label{eq:shift}
\textit{shift}(\mathbf{x}, \delta) = (x_{(0+\delta)\inmod{M}}, \dots, x_{(M-1+\delta)\inmod{M}})^\top,
\end{equation}
and $\textit{shift}(\mathbf{x}_{t-n}^{t}, \delta)$ denotes the transposition of each single time step vector \emph{before} concatenation and linearization.

The training procedure is then as follows: 
First, the mapping code $\mathbf{m}_{t+1}$ of an input/target pair is inferred as shown in Equation \ref{eq:gamap}. Then, $\mathbf{m}_{t+1}$ is used to reconstruct a \emph{transposed} version of the target, from an equally \emph{transposed} input (modifying Equation \ref{recony}) as
\begin{equation}\label{reconyshift}
\mathbf{\tilde{x}}'_{t+1} = \sigma_g(\mathbf{V}^\top (\mathbf{W}_0^\top\mathbf{W}_1^\top \mathbf{m}_t \cdot \mathbf{U}\textit{shift}(\mathbf{x}_{t-n}^{t},\delta))),
\end{equation}
with $\delta \in [-60,60]$ randomly chosen for each training batch.
Finally, we penalize the error between the reconstruction of the transposed target and the actual transposed target (i.e., employing Equation \ref{eq:cost}) as
\begin{equation}
\text{MSE} = \frac{1}{M} \left\| \textit{shift}(\mathbf{x}_{t+1},\delta) - \mathbf{\tilde{x}}'_{t+1} \right\|^2.
\end{equation}

This method amounts to both, a form of guided training and data augmentation.

\subsubsection{Training Details}
The architecture and training details of the GAE are as follows. In this paper, we use two models with differing n-gram lengths. The factor layer has $512$ units for n-gram length $n=16$, and $256$ units for $n=8$. Furthermore, for all models, there are $128$ neurons in the first mapping layer and $64$ neurons in the second mapping layer, i.e. the features we will be using throughout this paper for the alignment task are $64$-dimensional.

L2 weight regularization for weights $\mathbf{U}$ and $\mathbf{V}$ is applied, as well as sparsity regularization \cite{Lee:2007uz} on the topmost mapping layer. 
The deviation of the norms of the columns of both weight matrices $\mathbf{U}$ and $\mathbf{V}$ from their average norm is penalized. 
Furthermore, we restrict these norms to a maximum value.
We apply $50\%$ dropout on the input and no dropout on the target, as proposed in \cite{memisevic2011gradient}. 
The learning rate (1e-3) is gradually decremented to zero over 300 epochs of training.

\section{Alignment Algorithm}\label{sec:alignment_algorithm}

The goal of this paper is to give the reader a good intuition about the novel transposition-invariant features for audio alignment and focus on their properties, without being distracted by a complicated alignment algorithm. Thus, we use a simple multi-scale variant of the dynamic time warping (DTW) algorithm (see \cite{Mueller15_FMP_SPRINGER} for a detailed description of DTW) for the experiments throughout the paper, namely FastDTW\cite{salvador:ida:2007} with the radius parameter set to 50. We performed all experiments presented in this paper using the cityblock, Euclidean and cosine distance measures to compute distances between feature vectors. Because the choice of distance measure did not have a sizeable impact, we only report the results using the Euclidean distance. As FastDTW is a well-known and widely used algorithm, we refrain from describing the algorithm here in detail and refer the reader to the referenced works.

Obviously, a large number of more sophisticated alternatives to FastDTW exists. This includes methods based on hidden Markov and semi-Markov models \cite{Raphael99_AutomaticSegmentation_IEEE-TPAMI,orio:icmc:2001,nakamura:etal:2014}, conditional random fields \cite{JoderER11_AudioScoreMatching_TASLP}, general graphical models \cite{Raphael01_probabilistic_JCGS, Raphael04_align_ISMIR, Cont10_ScoreMusic_IEEE-TPAMI, maezawa:ismir:2014}, Monte Carlo sampling \cite{MontecchioC11_RealTimeSyncViaSequentialMonteCarlo_ICASSP,DuanP11_StateSpaceModelAlignment_ICASSP}, and extensions to DTW, e.g. multiple sequence alignment \cite{wang:ismir:2014} and integrated tempo models \cite{arzt:smc:2010}. We are confident that the presented features can also be employed successfully with these more sophisticated alignment schemes.

\section{Experiments on Piano Music}\label{sec:experiments_piano}

In this section we present a number of experiments, showcasing the strengths of the proposed features as well as their weaknesses. We will do this on piano music first, before moving on to more complex orchestral music in Section \ref{sec:experiments_orchestra}. For learning the features, a dataset consisting of 100 random piano pieces of the MAPS dataset \cite{emiya2010multipitch} (subset MUS) was used. As discussed in Section \ref{sec:feature_computation}, no annotations are needed, thus actually any available audio recording of piano music could be used. For the experiments, we trained two models, differing in the size of their local context: an 8-gram model and a 16-gram model (referred to as \emph{8G Piano} and \emph{16G Piano} in the remainder of the paper).

For the evaluation of audio-to-score alignment, a collection of annotated test data (pairs of scores and exactly aligned performances) is needed. We performed experiments on four datasets (see Table \ref{tab:data_set}).  \emph{CB} and \emph{CE} consist of 22 recordings of the Ballade Op. 38 No. 1 and the Etude Op. 10 No. 3 by Chopin \cite{vienna4x22}, \emph{MS} contains performances of the first movements of the piano sonatas KV279-284, KV330-333, KV457, KV475 and KV533 by Mozart \cite{widmer:aimag}, and \emph{RP} consists of three performances of the Prelude Op. 23 No. 5 by Rachmaninoff \cite{arzt:2016}. The scores are provided in the MIDI format. Their global tempo is set such that the score audio roughly matches the mean length of the given performances. The scores are then synthesised with the help of \emph{timidity}\footnote{\url{https://sourceforge.net/projects/timidity/}} and a publicly available sound font. The resulting audio files are used as score representations for the alignment experiments.

\begin{table}
\begin{center}
\begin{tabular}{clcc}
\toprule
\textbf{ID} & \textbf{Dataset} & \textbf{Files} & \textbf{Duration}\\ \midrule
\textbf{CE} & Chopin Etude  & 22 &  $\sim$ 30 min.\\ 
\textbf{CB} & Chopin Ballade & 22 & $\sim$ 48 min. \\ 
\textbf{MS} & Mozart Sonatas & 13 &  $\sim$ 85 min.\\ 
\textbf{RP} & Rachmaninoff Prelude & 3 & $\sim$ 12 min. \\ 
\bottomrule
\end{tabular}
\caption{The evaluation data set for the experiments on piano music (see text).}
\label{tab:data_set}
\end{center}
\end{table}

In the experiments, we use two types of evaluation measures. For each experiment, the 1\textsuperscript{st} quartile, the median, and the 3\textsuperscript{rd} quartile of the absolute errors at aligned reference points is given. We also report the percentage of reference points which have been aligned with errors smaller or equal 50 ms, and smaller or equal 250 ms  (similar to \cite{dixon:ismir:2005}).

\subsection{Experiment 1: Feature Configurations}

The first experiment compares the performance of the two feature configurations \emph{8G Piano} and \emph{16G Piano} on the piano evaluation set (see Table \ref{tab:exp1}). The differences between the two configurations are relatively small, although the 8-gram feature consistently works slightly better than the 16-gram features. The danger of using a larger local context is that different tempi can lead to very different contexts (e.g. faster tempi result in more notes contained in the local context), which in turn leads to different features, which is a problem for the matching process. We will return to this problem in a later experiment (see Section \ref{sec:exp5}). Because of space constraints, in the upcoming sections, we will only report the results for \emph{8G Piano}.

\begin{table}[]
	\centering
	\begin{tabular}{llcc}
	\toprule
		\textbf{Dataset}             & \textbf{Measure}         & \textbf{8G Piano}                   & \textbf{16G Piano}                                  \\ \midrule
		\multirow{5}{*}{\textbf{CB}} & 1\textsuperscript{st} Quartile     & \textbf{\SI{10}{\milli\second}}      & \SI{11}{\milli\second}   \\
		                    & Median            & \textbf{\SI{22}{\milli\second}}    & \SI{24}{\milli\second}  \\
		                    & 3\textsuperscript{rd} Quartile         & \textbf{\SI{39}{\milli\second}}  & \SI{45}{\milli\second}  \\
		                    & Error $\leq$ \SI{50}{\milli\second}        & \textbf{83\%}           & 79\%                                      \\
		                    & Error $\leq$ \SI{250}{\milli\second}                & {94\%}               & \textbf{95\%}                            \\ \midrule
		\multirow{5}{*}{\textbf{CE}} & 1\textsuperscript{st} Quartile       & \textbf{\SI{10}{\milli\second}}     & \SI{12}{\milli\second}\\
		                    & Median            & \textbf{\SI{21}{\milli\second}}     & \SI{25}{\milli\second}  \\
		                    & 3\textsuperscript{rd} Quartile        & \textbf{\SI{36}{\milli\second}}    & \SI{45}{\milli\second}  \\
		                    & Error $\leq$ \SI{50}{\milli\second}          & \textbf{87\%}           & 79\%                                          \\
		                    & Error $\leq$ \SI{250}{\milli\second}        & \textbf{96\%}             & 95\%                                       \\ \midrule
		\multirow{5}{*}{\textbf{MS}} & 1\textsuperscript{st} Quartile       & \textbf{\SI{6}{\milli\second}}      & \textbf{\SI{6}{\milli\second}}  \\
		                    & Median              & \textbf{\SI{13}{\milli\second}}    & \SI{14}{\milli\second} \\
		                    & 3\textsuperscript{rd} Quartile        & \textbf{\SI{25}{\milli\second}}   & \SI{26}{\milli\second}  \\
		                    & Error $\leq$ \SI{50}{\milli\second}        & {90\%}        & \textbf{91\%}                                       \\
		                    & Error $\leq$ \SI{250}{\milli\second}        & \textbf{100\%}          & \textbf{100\%}                                   \\ \midrule
		\multirow{5}{*}{\textbf{RP}} & 1\textsuperscript{st} Quartile    & \textbf{\SI{14}{\milli\second}}   & \SI{16}{\milli\second} \\
		                    & Median             & \textbf{\SI{34}{\milli\second}}     & \SI{40}{\milli\second} \\
		                    & 3\textsuperscript{rd} Quartile        & \textbf{\SI{90}{\milli\second}}    & \SI{92}{\milli\second}  \\
		                    & Error $\leq$ \SI{50}{\milli\second}                  & \textbf{63\%}            & 57\%                                    \\
		                    & Error $\leq$ \SI{250}{\milli\second}                               & 90\%           & \textbf{93\%}                 \\
		                    \bottomrule
	\end{tabular}
	\caption{Comparison of the 8-gram and the 16-gram feature models.}
		\label{tab:exp1}

\end{table}

\subsection{Experiment 2: Transposition-invariance}\label{sec:transp_invariance}

Next, we demonstrate that the learned features are actually invariant to transpositions. To do so, we transposed the score representations by -3, -2, -1, 0, +1, +2 and +3 semitones and tried to align the untransposed performances to these scores. The results for the \emph{8G Piano} features are shown in Table \ref{tab:exp2}. The results for all transpositions, including the untransposed scores, are very similar. Only minor fluctuations occur randomly.

In addition, we prepared a second, more challenging experiment. We manipulated the scores such that after every 30 seconds another transposition from the set $\{-3, -2, -1, +1, +2, +3\}$ is randomly applied. From each score, we created five such randomly changing score representations and tried to align the performances to these scores. The results are shown in the rightmost column of Table \ref{tab:exp2}. Again, there is no difference to the other results. Basically, the transpositions only lead to at most eight noisy feature vectors every time a new transposition is applied, which is not a problem for the alignment algorithm. We would also like to note that very few algorithms or features would be capable of solving this task (see \cite{MuellerC07_Transposition_ISMIR} for another option). Other methods that first try to globally identify the transposition and then use traditional methods for the alignment are clearly not applicable here.

\begin{table*}[]
	\centering
	\begin{tabular}{llcccccccc}
	\toprule
		& & \multicolumn{7}{c}{\textbf{Transposition in Semitones}} \\ \cmidrule(r){3-9}
		\textbf{Dataset}            & \textbf{Measure}                & \textbf{-3}                      & \textbf{-2}                      & \textbf{-1}                      & \textbf{0}                      & \textbf{1}                       & \textbf{2}                      & \textbf{3}     & \textbf{Rand. Transp.}                   \\ \midrule
		\multirow{5}{*}{\textbf{CB}} & 1\textsuperscript{st} Quartile           & \SI{10}{\milli\second}  & \SI{10}{\milli\second}  & \SI{11}{\milli\second}  & \SI{10}{\milli\second} & \SI{10}{\milli\second}  & \SI{11}{\milli\second} & \SI{10}{\milli\second} & \SI{10}{\milli\second}  \\
		                    & Median                 & \SI{22}{\milli\second}  & \SI{22}{\milli\second}  & \SI{23}{\milli\second}  & \SI{22}{\milli\second} & \SI{22}{\milli\second}  & \SI{23}{\milli\second} & \SI{22}{\milli\second} & \SI{22}{\milli\second} \\
		                    & 3\textsuperscript{rd} Quartile           & \SI{39}{\milli\second}  & \SI{39}{\milli\second}  & \SI{41}{\milli\second}  & \SI{39}{\milli\second} & \SI{40}{\milli\second}  & \SI{40}{\milli\second} & \SI{38}{\milli\second} & \SI{39}{\milli\second}  \\
		                    &Error $\leq$ \SI{50}{\milli\second}  & 84\%                  & 83\%                  & 81\%                  & 83\%                 & 82\%                  & 83\%                 & 84\%         & 83\%           \\
		                    & Error $\leq$ \SI{250}{\milli\second} & 95\%                  & 94\%                  & 94\%                  & 94\%                 & 93\%                  & 95\%                 & 95\%       & 95\%            \\ \midrule
		\multirow{5}{*}{\textbf{CE}} & 1\textsuperscript{st} Quartile           & \SI{10}{\milli\second}   & \SI{10}{\milli\second}  & \SI{8}{\milli\second}  & \SI{10}{\milli\second} & \SI{9}{\milli\second}   & \SI{9}{\milli\second}  & \SI{10}{\milli\second} & \SI{9}{\milli\second} \\
		                    & Median                 & \SI{21}{\milli\second}  & \SI{20}{\milli\second}  & \SI{18}{\milli\second}  & \SI{21}{\milli\second} & \SI{19}{\milli\second}  & \SI{18}{\milli\second} & \SI{20}{\milli\second} & \SI{19}{\milli\second}  \\
		                    & 3\textsuperscript{rd} Quartile           & \SI{37}{\milli\second}  & \SI{32}{\milli\second}  & \SI{30}{\milli\second}  & \SI{36}{\milli\second} & \SI{33}{\milli\second}  & \SI{32}{\milli\second} & \SI{33}{\milli\second} & \SI{32}{\milli\second} \\
		                    & Error $\leq$ \SI{50}{\milli\second}  & 83\%                  & 90\%                  & 91\%                  & 87\%                 & 89\%                  & 88\%                 & 90\%        & 90\%           \\
		                    & Error $\leq$ \SI{250}{\milli\second} & 93\%                  & 97\%                  & 98\%                  & 96\%                 & 97\%                  & 98\%                 & 97\%         & 97\%          \\ \midrule
		\multirow{5}{*}{\textbf{MS}} & 1\textsuperscript{st} Quartile           & \SI{6}{\milli\second}   & \SI{6}{\milli\second}   & \SI{6}{\milli\second}   & \SI{6}{\milli\second}  & \SI{6}{\milli\second}   & \SI{6}{\milli\second}  & \SI{6}{\milli\second} & \SI{6}{\milli\second}   \\
		                    & Median                 & \SI{13}{\milli\second}  & \SI{13}{\milli\second}  & \SI{13}{\milli\second}  & \SI{13}{\milli\second} & \SI{13}{\milli\second}  & \SI{14}{\milli\second} & \SI{13}{\milli\second} & \SI{13}{\milli\second}   \\
		                    & 3\textsuperscript{rd} Quartile           & \SI{24}{\milli\second}  & \SI{24}{\milli\second}  & \SI{24}{\milli\second}  & \SI{25}{\milli\second} & \SI{25}{\milli\second}  & \SI{26}{\milli\second} & \SI{25}{\milli\second} & \SI{25}{\milli\second}  \\
		                    & Error $\leq$ \SI{50}{\milli\second}  & 91\%                  & 91\%                  & 92\%                  & 90\%                 & 91\%                  & 90\%                 & 90\%         & 91\%         \\
		                    & Error $\leq$ \SI{250}{\milli\second} & 100\%                 & 100\%                 & 100\%                 & 100\%                & 100\%                 & 100\%                & 100\%     & 100\%            \\ \midrule
		\multirow{5}{*}{\textbf{RP}} & 1\textsuperscript{st} Quartile           & \SI{17}{\milli\second}  & \SI{16}{\milli\second}  & \SI{15}{\milli\second}  & \SI{14}{\milli\second} & \SI{13}{\milli\second}  & \SI{12}{\milli\second} & \SI{15}{\milli\second}  & \SI{14}{\milli\second}  \\
		                    & Median                 & \SI{45}{\milli\second}  & \SI{44}{\milli\second}  & \SI{36}{\milli\second}  & \SI{34}{\milli\second} & \SI{35}{\milli\second}  & \SI{31}{\milli\second} & \SI{34}{\milli\second} & \SI{37}{\milli\second}  \\
		                    & 3\textsuperscript{rd} Quartile           & \SI{136}{\milli\second} & \SI{151}{\milli\second} & \SI{106}{\milli\second} & \SI{90}{\milli\second} & \SI{130}{\milli\second} & \SI{90}{\milli\second} & \SI{103}{\milli\second}  & \SI{122}{\milli\second} \\
		                    & Error $\leq$ \SI{50}{\milli\second}  & 53\%                  & 53\%                  & 60\%                  & 63\%                 & 59\%                  & 64\%                 & 60\%        & 58\%                \\
		                    & Error $\leq$ \SI{250}{\milli\second} & 84\%                  & 83\%                  & 88\%                  & 90\%                 & 85\%                  & 90\%                 & 89\%           & 86\%             \\ 
		                    \bottomrule
	\end{tabular}
	\caption{Results for 8-gram piano features on transposed versions of the scores (from -3 to +3 semitones). The rightmost column gives the results on scores with randomly changing transpositions after every 30 seconds (see Section \ref{sec:transp_invariance}).}
		\label{tab:exp2}

\end{table*}

\subsection{Experiment 3: Comparison to Chroma Features}

It is now time to compare the \emph{8G Piano} features to well-established features for the task of music alignment in the normal, un-transposed alignment setting. To this end, we computed the \emph{chroma\_cqt} features\footnote{We also tried the \emph{CENS} features, which are a variation of chroma features, but as they consistently performed worse than the \emph{Chroma} features, we are not reporting the results here.} (henceforth referred to as \emph{Chroma}) as provided by \emph{librosa}\footnote{Version 0.6,  DOI:10.5281/zenodo.1174893} \cite{mcfee:pis:2015} (with standard parameters except for the normalisation parameter, which we set to 1; the hop size is roughly \SI{20}{\milli\second}), and aligned the performances to the scores. The results are shown in Table \ref{tab:exp4}. On this dataset, the proposed \emph{transposition-invariant} features consistently outperform the well-established \emph{Chroma} features, which are based on \emph{absolute pitches}. To summarise, so far the proposed features show state-of-the-art performance on the standard alignment task, while additionally being able to align transposed sequences to each other with no additional error.

\begin{table}[]
	\centering
	\begin{tabular}{llccc}
	\toprule
		\textbf{DS}                  & \textbf{Measure}                & \textbf{Chroma}                                    & \textbf{8G Piano}                \\ \midrule
		\multirow{5}{*}{\textbf{CB}} & 1\textsuperscript{st} Quartile           & \SI{15}{\milli\second}   & \textbf{\SI{10}{\milli\second}} \\
		                    & Median                 & \SI{34}{\milli\second}   & \textbf{\SI{22}{\milli\second}} \\
		                    & 3\textsuperscript{rd} Quartile           & \SI{80}{\milli\second}  & \textbf{\SI{39}{\milli\second}} \\
		                    & Error $\leq$ \SI{50}{\milli\second}  & 64\%                               & \textbf{83\%}                 \\
		                    & Error $\leq$ \SI{250}{\milli\second} & 85\%                            & \textbf{94\%}                 \\ \midrule
		\multirow{5}{*}{\textbf{CE}} & 1\textsuperscript{st} Quartile           & \SI{13}{\milli\second}   & \textbf{\SI{10}{\milli\second}} \\
		                    & Median                 & \SI{29}{\milli\second}  & \textbf{\SI{21}{\milli\second}} \\
		                    & 3\textsuperscript{rd} Quartile           & \SI{56}{\milli\second}  & \textbf{\SI{36}{\milli\second}} \\
		                    & Error $\leq$ \SI{50}{\milli\second}  & 71\%                                & \textbf{87\%}                 \\
		                    & Error $\leq$ \SI{250}{\milli\second} & 94\%                                  & \textbf{96\%}                 \\ \midrule
		\multirow{5}{*}{\textbf{MS}} & 1\textsuperscript{st} Quartile           & \SI{7}{\milli\second}   & \textbf{\SI{6}{\milli\second}}  \\
		                    & Median                 & \SI{16}{\milli\second}  & \textbf{\SI{13}{\milli\second}} \\
		                    & 3\textsuperscript{rd} Quartile           & \SI{31}{\milli\second}   & \textbf{\SI{25}{\milli\second}} \\
		                    & Error $\leq$ \SI{50}{\milli\second} & 85\%                                  & \textbf{90\%}                 \\
		                    & Error $\leq$ \SI{250}{\milli\second} & 98\%                                & \textbf{100\%}                \\ \midrule
		\multirow{5}{*}{\textbf{RP}} & 1\textsuperscript{st} Quartile           & \SI{17}{\milli\second}   & \textbf{\SI{14}{\milli\second}} \\
		                    & Median                 & \SI{43}{\milli\second}   & \textbf{\SI{34}{\milli\second}} \\
		                    & 3\textsuperscript{rd} Quartile           & \SI{113}{\milli\second} & \textbf{\SI{90}{\milli\second}} \\
		                    & Error $\leq$ \SI{50}{\milli\second}  & 55\%                                 & \textbf{63\%}                 \\
		                    & Error $\leq$ \SI{250}{\milli\second} & \textbf{91\%}                                & $90\%$                 \\ 
		                    \bottomrule
	\end{tabular}
	\caption{Comparison of the transposition-invariant \emph{8G Piano} features to the \emph{Chroma} features on untransposed scores.}
		\label{tab:exp4}

\end{table}

\subsection{Experiment 4: Robustness to Tempo Variations}\label{sec:exp5}

Next, we have a closer look at the influence of different tempi on our features. As they are based on a fixed local context (a fixed number of frames), the tempo plays an important role in their computation. For example, if the tempo doubles, this means that musically speaking the local context is twice as large as at the normal tempo and additional notes might be included in this context, which would not be part of the local context in the case of the normal tempo. To test the influence of tempo differences, we created score representations using different tempi and aligned the unchanged performances to them. Table \ref{tab:exp5} summarises the results for the \emph{Chroma} and the \emph{8G Piano} features on scores synthesised with the base tempo, as well as with $\frac{2}{3}$-times and $\frac{4}{3}$-times the base tempo. Unsurprisingly, tempo in general influences the alignment results. However, while the \emph{Chroma} features are much more robust to differences in tempo between the sequences to be aligned, the \emph{8G Piano} features struggle in this experiment. We repeated the experiment with more extreme tempo changes, which confirmed this trend. While with the \emph{Chroma} features it is possible to more or less align sequences with tempo differences of a factor of three, the transposition-invariant features fail in these cases.

\begin{table*}[]
	\centering
	\begin{tabular}{llcccccc}
	\toprule
	& &  \multicolumn{3}{c}{\textbf{Chroma}} & \multicolumn{3}{c}{\textbf{8G Piano}}  \\ \cmidrule(r){3-5} \cmidrule(r){6-8} 
		\textbf{DS}                  & \textbf{Measure}                & \textbf{$\mathbf{\frac{2}{3}}$  Tempo}                & \textbf{Base Tempo}                    &\textbf{$\mathbf{\frac{4}{3}}$  Tempo}       & \textbf{$\mathbf{\frac{2}{3}}$  Tempo}                   & \textbf{Base Tempo}                    &\textbf{$\mathbf{\frac{4}{3}}$  Tempo}             \\ \midrule
		\multirow{5}{*}{\textbf{CB}} & 1\textsuperscript{st} Quartile           & \SI{19}{\milli\second} & \SI{15}{\milli\second} & \SI{13}{\milli\second} & \SI{24}{\milli\second} & \SI{10}{\milli\second} & \SI{32}{\milli\second}   \\
		                    & Median                 & \SI{43}{\milli\second}  & \SI{34}{\milli\second} & \SI{29}{\milli\second} & \SI{63}{\milli\second} & \SI{22}{\milli\second} & \SI{120}{\milli\second} \\
		                    & 3\textsuperscript{rd} Quartile           & \SI{137}{\milli\second}  & \SI{80}{\milli\second} & \SI{66}{\milli\second} & \SI{116}{\milli\second} & \SI{39}{\milli\second} & \SI{205}{\milli\second}  \\
		                    & Error $\leq$ \SI{50}{\milli\second}  & 54\%  & 64\%  & 67\%  & 47\%  & 83\%  & 33\%                             \\
		                    & Error $\leq$ \SI{250}{\milli\second}   & 82\%  & 85\%  & 85\%  & 87\%  & 94\%  & 84\%                       \\ \midrule
		\multirow{5}{*}{\textbf{CE}} & 1\textsuperscript{st} Quartile           & \SI{14}{\milli\second}   & \SI{13}{\milli\second}  & \SI{12}{\milli\second}  & \SI{27}{\milli\second}  & \SI{10}{\milli\second}  & \SI{26}{\milli\second}  \\
		                    & Median                 & \SI{30}{\milli\second}   & \SI{29}{\milli\second}  & \SI{25}{\milli\second}  & \SI{70}{\milli\second}  & \SI{21}{\milli\second}  & \SI{83}{\milli\second}   \\
		                    & 3\textsuperscript{rd} Quartile           & \SI{65}{\milli\second}   & \SI{56}{\milli\second}  & \SI{53}{\milli\second}  & \SI{116}{\milli\second}  & \SI{36}{\milli\second}  & \SI{176}{\milli\second}   \\
		                    & Error $\leq$ \SI{50}{\milli\second} & 69\%  & 71\%  & 73\%  & 40\%  & 87\%  & 38\%                               \\
		                    & Error $\leq$ \SI{250}{\milli\second} & 90\%  & 94\%  & 94\%  & 93\%  & 96\%  & 80\%                          \\ \midrule
		\multirow{5}{*}{\textbf{MS}} & 1\textsuperscript{st} Quartile         & \SI{8}{\milli\second}   & \SI{7}{\milli\second}  & \SI{9}{\milli\second}  & \SI{7}{\milli\second}  & \SI{6}{\milli\second}  & \SI{9}{\milli\second}  \\
		                    & Median                 & \SI{18}{\milli\second}   & \SI{16}{\milli\second}  & \SI{20}{\milli\second}  & \SI{16}{\milli\second}  & \SI{13}{\milli\second}  & \SI{21}{\milli\second} \\
		                    & 3\textsuperscript{rd} Quartile           & \SI{42}{\milli\second}   & \SI{31}{\milli\second}  & \SI{49}{\milli\second}  & \SI{33}{\milli\second}  & \SI{25}{\milli\second}  & \SI{52}{\milli\second}  \\
		                    & Error $\leq$ \SI{50}{\milli\second} & 79\%  & 85\%  & 75\%  & 84\%  & 90\%  & 74\%                             \\
		                    & Error $\leq$ \SI{250}{\milli\second} & 98\%  & 98\%  & 97\%  & 99\%  & 100\%  & 98\%                        \\ \midrule
		\multirow{5}{*}{\textbf{RP}} & 1\textsuperscript{st} Quartile       & \SI{18}{\milli\second}   & \SI{17}{\milli\second}  & \SI{20}{\milli\second}  & \SI{22}{\milli\second}  & \SI{14}{\milli\second}  & \SI{30}{\milli\second} \\
		                    & Median                 & \SI{44}{\milli\second}   & \SI{43}{\milli\second}  & \SI{58}{\milli\second}  & \SI{69}{\milli\second}  & \SI{34}{\milli\second}  & \SI{86}{\milli\second} \\
		                    & 3\textsuperscript{rd} Quartile           & \SI{116}{\milli\second}   & \SI{113}{\milli\second}  & \SI{141}{\milli\second}  & \SI{184}{\milli\second}  & \SI{90}{\milli\second}  & \SI{202}{\milli\second}  \\
		                    & Error $\leq$ \SI{50}{\milli\second}  & 53\%  & 55\%  & 56\%  & 43\%  & 63\%  & 37\%                           \\
		                    & Error $\leq$ \SI{250}{\milli\second} & 92\%  & 91\%  & 87\%  & 82\%  & 90\%  & 85\%                         \\ 
		                    \bottomrule
	\end{tabular}
	\caption{Comparison of Chroma and 8G Piano features for alignments to scores in different tempi (see Section \ref{sec:exp5}).}
		\label{tab:exp5}

\end{table*}

\section{First Experiments on Orchestral Music}\label{sec:experiments_orchestra}

\begin{table}[]
	\centering
	\begin{tabular}{llcccc}
	\toprule
		\textbf{DS}                  & \textbf{Measure}                & \textbf{Chroma}                             & \textbf{8G Piano}       & \textbf{8G Orch}            \\ \midrule
		\multirow{5}{*}{\textbf{B3}} & 1\textsuperscript{st} Quartile           & \textbf{\SI{20}{\milli\second}}   & \SI{25}{\milli\second} & \SI{22}{\milli\second}  \\
		                    & Median                 & \textbf{\SI{48}{\milli\second}}  &\SI{54}{\milli\second}  & \SI{49}{\milli\second} \\
		                    & 3\textsuperscript{rd} Quartile           & {\SI{108}{\milli\second}} &\textbf{\SI{104}{\milli\second}} & \SI{111}{\milli\second}  \\
		                    & Err. $\leq$ \SI{50}{\milli\second}   & \textbf{52\%}                &    47\%         & 51\%                 \\
		                    & Err. $\leq$ \SI{250}{\milli\second} & {88\%}             &      \textbf{90\%}            & 89\%                 \\ \midrule
		\multirow{5}{*}{\textbf{M4}} & 1\textsuperscript{st} Quartile           & \textbf{\SI{46}{\milli\second}}  &  \SI{50}{\milli\second}  & \SI{57}{\milli\second}  \\
		                    & Median                 & \textbf{\SI{110}{\milli\second}} &  \SI{129}{\milli\second}  & \SI{142}{\milli\second}  \\
		                    & 3\textsuperscript{rd} Quartile           & \textbf{\SI{278}{\milli\second}} &  \SI{477}{\milli\second}  & \SI{535}{\milli\second}  \\
		                    & Err. $\leq$ \SI{50}{\milli\second}   & \textbf{27\%}                     &     25\%       & 23\%                 \\
		                    & Err. $\leq$ \SI{250}{\milli\second} & \textbf{73\%}                  &      66\%      & 62\%                 \\ 
		                    \bottomrule
	\end{tabular}
	\caption{Comparison of the transposition-invariant and chroma features on orchestral music (see Section~\ref{sec:experiments_orchestra}).}
		\label{tab:exp_orchestra}
\end{table}

In addition to the promising results on piano music, we also present first experiments on orchestral music. To this end, we trained an additional model on recordings of symphonic music (seven full commercial recordings of symphonies by Beethoven, Brahms, Bruckner, Berlioz and Strauss), which will be referred to in the following as \emph{8G Orch}. For comparison, we also evaluated the model from the previous section (\emph{8G Piano}) and the \emph{Chroma} features on the evaluation data. The evaluation data consists of two recordings of classical symphonies: the 3\textsuperscript{rd} symphony by Beethoven (B3) and the 4\textsuperscript{th} symphony by Mahler (M4). Both have been manually annotated at the downbeat level.

In this alignment experiment, the \emph{Chroma} features outperform both the \emph{8G Piano} and the \emph{8G Orch} features, especially on the symphony by Mahler (see Table \ref{tab:exp_orchestra}). We mainly contribute this to the fact that these rather long recordings contain a number of sections with different tempi, which is not reflected in the score representations. As has been established in Section \ref{sec:exp5}, the transposition-invariant features struggle in these cases. Still, we will have to further investigate the use of these features for orchestral music.

It is interesting to note that \emph{8G Piano} gives slightly better results than \emph{8G Orch}, even though this dataset solely consists of orchestral music. It turns out that the learned features are very general and can be readily applied to different instruments. We also tried to overfit on the test data, i.e., we trained a feature model using the audio files we would later use for the alignment experiments. Even this approach only led to fractionally better results.


\section{Conclusions}\label{sec:conclusion}

In this paper, we reported on audio-to-score alignment experiments with novel transposition-invariant features. We have shown that the features are indeed fully invariant to transpositions and in many settings can outperform the quasi-standard features for this task, namely chroma-based features. On the other hand, we also demonstrated the weaknesses of the transposition-invariant features, especially their fragility regarding different tempi, which is a serious limitation in the context of alignment tasks.

In the future, we will study this weakness in depth and will try to alleviate this problem. Ideas include further experiments with different n-gram lengths, the adoption of alignment schemes including tempo models which iteratively adapt the local tempi of the representations, and to try to include tempo-invariance as an additional goal in the learning process of the features.

\section{Acknowledgements}

This research has received funding from the European Research Council
(ERC) under the European Union's Horizon 2020 research and innovation
programme (ERC grant agreement No. 670035, project CON ESPRESSIONE). 

\bibliography{bibliography}

\end{document}